\documentclass[
    ,final            
  ,numberedheadings 
  ]
  {aipproc}

\layoutstyle{8x11double}

\begin{document}

\title{The Cross-Calibration of Swift-BAT and Fermi-GBM via Correlative Spectral Analysis of GRBs}


\author{Michael Stamatikos on behalf of the BAT/GBM Inter-calibration Team}{
  address={Center for Cosmology and Astro-Particle Physics (CCAPP) Fellow/Department of Physics, \\The Ohio State University, 191 West Woodruff Avenue, Columbus, OH 43210, USA
\\Michael.Stamatikos-1@nasa.gov}}

\begin{abstract}
We report on recent inter-calibration studies featuring Swift's Burst Alert Telescope (BAT) and Fermi's Gamma-ray Burst Monitor (GBM) based upon correlated observations of GRBs 080804 and 080810, via their resultant joint spectral analysis. Swift's intrinsic multi-wavelength instrumentation and dynamical response complement Fermi's superior energy range. The addition of BAT's spectral response will (i) facilitate in-orbit GBM detector response calibration, (ii) augment Fermi's low energy sensitivity, (iii) enable ground-based follow-up efforts of Fermi GRBs, and (iv) help identify a subset of GRBs discovered via off-line GBM data analysis, for an annual estimate of $\sim30$ GRBs. The synergy of BAT and GBM augments previous successful joint spectral fit efforts by enabling the study of peak photon energies (Epeak), while leveraging the over eleven energy decades afforded by Fermi's Large Area Telescope (LAT), in conjunction with Swift's X-Ray (XRT) and Ultraviolet-Optical (UVOT) Telescopes, for an unprecedented probe of broad-band spectral and temporal evolution, throughout their contemporaneous orbital tenure over the next decade.
\end{abstract}

\maketitle


\section{OBJECTIVES \& MOTIVATION}

The Swift MIDEX explorer mission \cite{Gehrels:2007}, comprised of the wide-field ($\sim1.4$ sr, half-coded) hard X-ray (15-150 keV) Burst Alert Telescope (BAT), and the narrow-field (0.2-10 keV) X-Ray (XRT) and (170-600 nm) Ultraviolet-Optical (UVOT) Telescopes, has revolutionized our understanding of gamma-ray bursts (GRBs) since its launch in late 2004. The intrinsic multi-wavelength instrumentation, coupled with a rapid ($<\sim$100 seconds) autonomous slew capability, has ushered in an unprecedented era of source localization precision $\left(<\sim1^{\prime}-4^{\prime}\right)$ that is disseminated in real-time ($\sim10$ seconds) via the GRB Coordinate Network (GCN), and spear-heads multi-wavelength international observational campaigns. Swift's unique dynamic response, in conjunction with correlative ground-based follow-up efforts, has resulted in more than $\sim133$ redshifts (z). Furthermore, Swift's X-ray sensitivity has revealed an afterglow paradigm as well as extended soft emission in some short GRBs. With funding recommendations until $\sim2010$ and an estimated orbital lifetime of $\sim15$ years, Swift will most likely operate until $\sim2017$.

The Fermi Gamma-ray Large Area Space Telescope \cite{Band:2009} launched on June 11, 2008 and has an anticipated mission lifetime of $\sim5$ and a goal of $\sim10$ years, respectively, taking Fermi into $\sim2018$. A goal of Fermi, which is comprised of the ($<$20 MeV to $>$300 GeV) Large Area Telescope (LAT) and the (10 keV - 30 MeV) Gamma-ray Burst Monitor (GBM), is to study transient gamma-ray sources, while a direct GBM experimental objective is to identify and study GRBs. Common scientific interest between Fermi and Swift provides strong motivation for using Fermi-GBM and Swift-BAT data to conduct a cross-calibration via correlative observations of GRBs, resulting in joint spectral fits, thus enabling the analysis of multi-wavelength spectral and temporal evolution.

\begin{figure}
  \includegraphics[height=.3\textheight,width=0.48\textwidth]{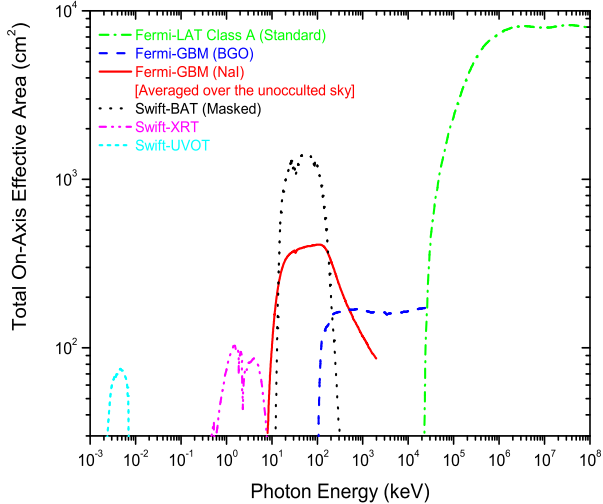}
  \caption{Effective areas for Swift-BAT/XRT/UVOT and Fermi-GBM/LAT. Exceptional bursts may be detected over 11 energy decades facilitating broad-band temporal and spectral evolution studies, as has been realized with GRB 090510.}
\label{Overlapped_Effective_Areas}
\end{figure}

The GBM, consisting of 12 NaI (10-1000 keV) and 2 BGO (0.15-25 MeV) detectors, monitors $\sim8$ steradians of the sky, and, in concert with LAT, enables Fermi to continuously span 7 energy decades, but not at the same sensitivity. As illustrated in Figure~\ref{Overlapped_Effective_Areas}, Fermi's effective area drops by over $\sim1.5$ orders of magnitude from LAT (GeV) to GBM-NaI (keV) energies, while the (masked) BAT ($\sim$20-100 keV) low energy effective area surpasses GBM-NaI's by over a factor of $\sim3$. Furthermore, although Swift has detected $\sim400$ GRBs, the majority of Epeak (characteristic photon energy) values lie beyond BAT's canonical energy range. Thus, correlated Swift-BAT/Fermi-GBM GRB observations would simultaneously augment Fermi's low energy response while increasing the number of Epeaks for BAT GRBs also observed by GBM. Additionally, since Swift's high fidelity localization precision surpasses GBM's by over $\sim2-3$ orders of magnitude, joint fits help guide selection for GRB follow-up campaigns. In this manner, we expect that $\sim35\%$ of bursts in the joint BAT-GBM analysis would be accompanied by panchromatic ground-based follow-up observations resulting in spectroscopic redshift determinations and host galaxy identifications. Lastly, cross-calibration with Swift-BAT would complement in-orbit calibration efforts of GBM's detector response matrices (DRMs), such as planned Earth occultation techniques, as was done with other high energy detectors such as Suzaku-WAM, resulting in a more accurate spectral characterization of GRB parameters.

\section{METHODOLOGY OF JOINT FITS}

The annual GRB trigger rate for BAT is $\sim100\pm10$ and the pre-launch estimate for GBM was $\sim200\pm20$. However, since GBM triggering was enabled on July 11, 2008, 69 GRBs were detected within 100 days corresponding to $\sim$250 GRBs per year. Assuming isotropic GRB spatial distribution with uncorrelated fields of view (FOVs) over homogenous sky coverage, we expected the pre-launch number of BAT GRBs within GBM's FOV $\left(n_{GBM}^{BAT}\right)$ and the number of GBM GRBs within BAT's FOV $\left(n_{BAT}^{GBM}\right)$ to be $\sim64\pm10$ and $\sim21\pm16$, respectively. Due to BAT's superior sensitivity, we assumed that all $n_{BAT}^{GBM}$ would trigger Swift. Hence, $n_{BAT}^{GBM}$ served as a conservative lower limit for the annual number of GRBs anticipated for BAT-GBM cross-calibration $\left(n_{GRB}^{XCal}\right)$. Based upon the implicit degeneracy, we expected the number of $n_{GBM}^{BAT}$ failing to trigger GBM $\left(n_{GBM}^{BAT\prime}\right)$ to be $\sim43\pm19$, i.e. we expected only $\sim33\%\pm26\%$ of $n_{GBM}^{BAT}$ would trigger GBM-NaI  on-board. Since the GBM count rate data are continuously down-linked in 8 energy channels at 256 ms temporal resolution and in 128 energy channels at 4 s temporal resolution, its possible to detect additional non-triggered GRBs $\left(n_{GBM}^{\prime}\right)$ on the ground.  Hence, GBM spectra may still be available in the absence of an on-board Fermi trigger. However, $n_{GBM}^{\prime}$ would suffer from reduced temporal and spectral resolution, with backgrounds that will be difficult to quantify for low peak flux, long duration GRBs. Selection effects, such as detector composition and long accumulation timescales, bias BAT towards long, soft GRBs with lower Epeak. Consequently, BAT GRBs comprise a separate statistical class, as is demonstrated by their fluence and redshift distributions, from classical Burst and Transient Source Experiment (BATSE) GRBs, which have been used to estimate the GBM on-board detection rate due to their similarities.

Hence, BAT can facilitate GBM off-line GRB identification efforts since a subset of $n_{GBM}^{\prime}$ would be populated by $n_{GBM}^{BAT\prime}$. We have since discovered our first $n_{GBM}^{\prime}$ via Swift. GRB 080906A was a posteriori identified in GBM data within $\sim9^{\circ}$ of the BAT localization, which is consistent with the expected GBM localization accuracy for weak bursts. Our $\texttt{GLASTspec}$\footnote{\texttt{GLASTSpec} is a Fermi web interface for \texttt{XSPEC} \url{(http://heasarc.gsfc.nasa.gov/webspec/GLASTspec.html)}.} simulations \cite{Stamatikos:2008} have illustrated that the estimated number of non-triggered BAT GRBs in GBM's FOV that would render good quality spectra in NaI as $n_{GBM}^{BAT\prime\prime}\sim11\pm6$. Consequently, we estimated that $n_{GRB}^{XCal}\approx n_{BAT}^{GBM}+n_{GBM}^{BAT\prime\prime}\sim32\pm17$, which is similar to the overlap observed with other high energy detectors. We noted that these were conservative estimates, tantamount to a few BAT-GBM GRB cross-calibrations per month, which could be improved by correlating BAT-GBM FOVs via maximizing the overlap of their respective sky pointing directions. Hence, although the GBM rate is $\sim25\%$ higher than pre-launch estimates, we are pleased to report that to date our joint observations are well within our initial predictions \cite{Stamatikos:2008}.

Since it is still relatively early in the Fermi mission, the GBM instrument has had in-flight data for only a short period of time. The systematic errors associated with GBM spectral analysis are currently large and unknown resulting in several caveats\footnote{See the Fermi Science Support Center (\url{http://fermi.gsfc.nasa.gov/ssc/data/analysis/GBM_caveats.html}).}. In particular, GBM's DRM is a function of time in the context of Fermi's differential orientation, which may be due to either (i) survey mode rocking ($\sim4^{\circ}$/min) or (ii) autonomous repointing (slews), both of which may greatly effect the spectral analysis of long GRBs. These slewing effects are exacerbated in concert with Swift, which spans $\Delta\theta\sim50^{\circ}$ within $<\sim75$ seconds. Hence, cross-calibration with BAT will aid in the evolution of GBM DRMs by improving the quality and confidence of GBM spectral fits and localization performance, thus facilitating the development of solutions to the above known issues.

\section{GRB CASE STUDIES: 080804 \& 080810}

\begin{figure}
  \includegraphics[height=0.26\textheight,width=0.48\textwidth]{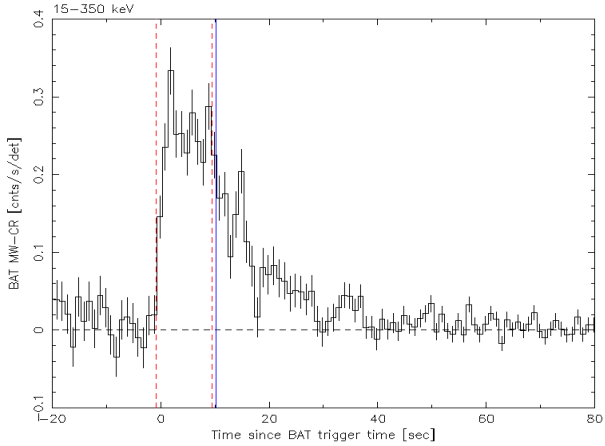}
  \caption{The Swift-BAT mask-weighted (15-350 keV) light curve for GRB 080804, with T$_{0}$ = 23:20:14 UTC. The solid (blue) line indicates Swift slew-time, while the dashed (red) lines indicate the joint fit interval. }
\label{GRB080804_LC}
\end{figure}

\begin{figure}
  \includegraphics[height=0.35\textheight,width=0.35\textwidth,angle=-90]{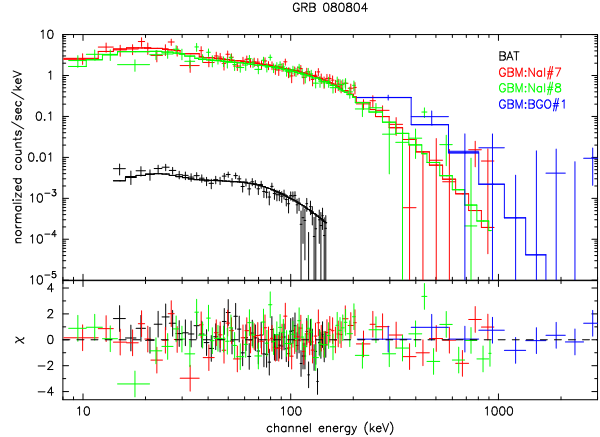}
  \caption{Joint Swift-BAT/Fermi-GBM spectral energy fit of GRB 080804, for a Comptonized model with parameters of $\alpha=0.61^{+0.12}_{-0.13}$ and Epeak $225^{+40}_{-30}$ keV ($\chi^2/$dof$\sim$1.14).}
\label{GRB080804}
\end{figure}

Initial case studies using GRBs 080804 and 080810 explored the treatment of slew intervals, while comparing the results of independent fitting software packages \texttt{XSPEC} and \texttt{RMFIT}, utilized by the BAT and GBM teams, respectively. For both cases, the Swift spectrum alone would have resulted in simple power law best fits, since the Epeak values were well beyond the BAT energy band pass (see Figure~\ref{Overlapped_Effective_Areas}), thus the GBM data were essential for their determination, as anticipated.

\textit{\textbf{GRB 080804}} was at a redshift of $\sim2.2$ and lasted for $\sim30$ sec, as illustrated in Figure~\ref{GRB080804_LC}. The integrated event spectrum was fitted with several functions, which included a (i) single power law, (ii) smoothly joined broken power law (Band function) and (iii) Comptonized model (Cut-off Power Law); the last of which provided the best fit. All three data sets, comprised of two GBM-NaI detectors, one GBM-BGO detector and Swift-BAT, agree extremely well, as can be seen in Figure~\ref{GRB080804}, which illustrates our first joint spectral fit results based upon BAT-GBM data for GRB $080804$.

\begin{figure}
  \includegraphics[height=.35\textheight,width=0.35\textwidth,angle=-90]{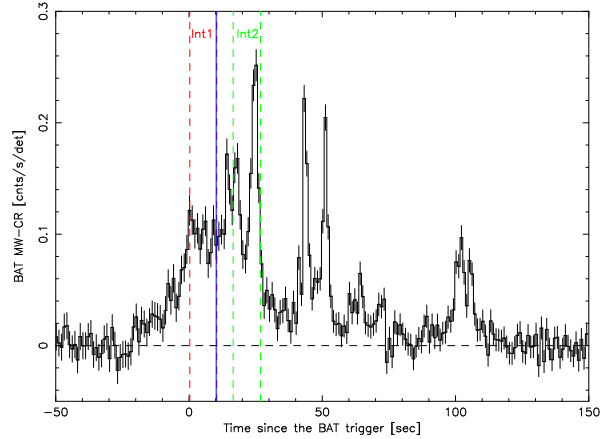}
  \caption{The Swift-BAT mask-weighted (15-350 keV) light curve of GRB 080810, with T$_{0}$=13:10:12.3 UTC. The solid (blue) and dashed (red/green) lines indicate Swift slew-time and the first/second joint fit intervals, respectively.}
\label{GRB080810_LC}
\end{figure}

\begin{figure}
  \includegraphics[height=.35\textheight,width=0.35\textwidth,angle=-90]{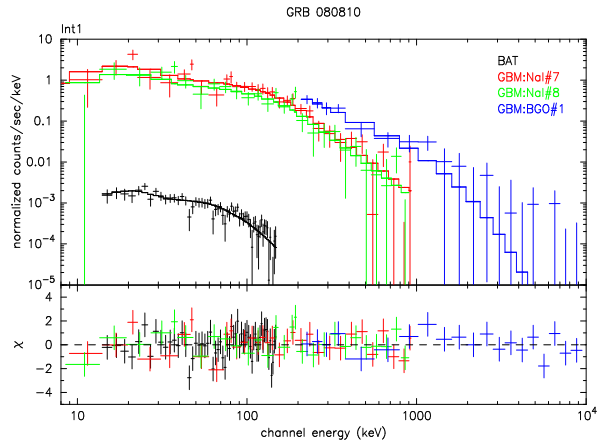}
  \caption{\small{Joint Swift-BAT/Fermi-GBM fit for the first interval of GRB 080810. A Comptonized model resulted in best fit parameters $\alpha=0.94^{+0.13}_{-0.15}$ and Epeak$=674^{+493}_{-237}$ keV ($\chi^2/$dof$\sim$1.33}), in agreement with the second interval fit.}
\label{GRB080810_1}
\end{figure}

\begin{figure}
  \includegraphics[height=.35\textheight,width=0.35\textwidth,angle=-90]{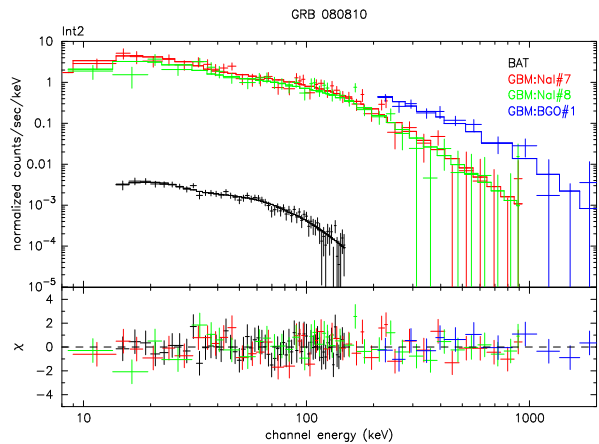}
  \caption{\small{Joint Swift-BAT/Fermi-GBM fit for the second interval of GRB 080810, whose relative brightness better constrains Epeak. A Comptonized model resulted in best fit parameters $\alpha=1.15^{+0.09}_{-0.10}$ and Epeak$=406^{+189}_{-106}$ keV ($\chi^2/$dof$\sim$1.15}).}
\label{GRB080810_2}
\end{figure}

\textit{\textbf{GRB 080810}} was a bright, multi-structured event at z = 3.35 that spanned over 100 sec, as illustrated by the Swift-BAT light curve in Figure~\ref{GRB080810_LC}. Swift autonomously slewed within 10 seconds, which complicates the DRM used in joint spectral fits with Fermi-GBM. In order to demonstrate this effect, integrated spectroscopy was performed for two sub-intervals. The first and second intervals spanned from $T_{GBM}+0<\sim t<\sim T_{GBM}+10$ and $T_{GBM}+16<\sim t<\sim T_{GBM}+26$ seconds, respectively. Two independent joint spectral fits for the first and second intervals were performed using a trinity of data from the BAT, NaI and BGO detectors, which were both best fit to a Comptonized Model, as illustrated in Figures~\ref{GRB080810_1} and \ref{GRB080810_2}, respectively. Although both Epeak values were consistent within the uncertainties, it was better constrained in the second interval where the burst was brighter. Independently, a time averaged joint BAT/Konus-Wind fit from $T_{BAT}-19<\sim t<\sim T_{BAT}+61$ seconds was made to a Comptonized model resulting in best fit parameters $\alpha=1.2\pm0.1$ and Epeak$=550^{+860}_{-230}$ keV ($\chi^2/$dof$\sim$0.71) \cite{Sakamoto:2008}. Although these results are not directly comparable due to differences in time interval and detector band pass, the general agreement is promising. Thus far, the normalization between BAT and GBM detectors appears to fluctuate about unity at a level of $\sim10-20\%$. However, bright events, such as GRB 090618, are necessary to analyze the DRM systematic trend.

\section{SCIENTIFIC IMPACT \& OUTLOOK}

Previous studies of joint fit GRB data sets \cite{Krimm:2006} enhanced our understanding of burst parameter classifications, explored GRB emission geometry, and tested the viability of various redshift estimation methods. Increasing the number of Swift Epeak values would help test the Sakamoto relation \cite{Sakamoto:2009}, which estimates the Epeak from the BAT photon index derived from a simple power-law fit. In addition, a more accurate normalization between Fermi prompt emission and Swift afterglow spectra will facilitate the determination of GRB energy budgets.

Via the Fermi-LAT, a natural extension into the GeV regime is possible for exceptional bursts. LAT has triggered on 9 GRBs, with ($68\%$) on-ground localization error radii\footnote{See Fermi Science Support Center LAT GRB lookup table (\url{http://fermi.gsfc.nasa.gov/ssc/dev/jd/grb_table/}).} of $\sim5^{\prime}<\sigma_{R}<\sim1^{\circ}$ ($\overline{\sigma_{R}}\sim18^{\prime}$). Bright LAT GRBs are prime candidates for discovery when coupled to redshift measurements, since they are expected to have on-board $\sigma_{R}<\sim20^{\prime}$ \cite{Kuehn:2007}, hence suitable for rapid target of opportunity follow-up via XRT and UVOT, Swift's narrow-field instruments (NFI), as has been done for GRBs 080916C and 090328. Conversely, if a GRB triggers both LAT and BAT, e.g. GRB 090510, chances favor subsequent detection(s) by Swift NFI's, since $\sim95\%$ of BAT GRBs are observed with XRT ($T<\sim200$ ksec) and $\sim60\%$ have accompanied optical measurements from a combination of UVOT and ground-based observations.

Swift-BAT/Fermi-GBM cross-calibration simultaneously provides performance diagnostics as well as enhanced discovery potential for Fermi science operations. In addition to our subset of over $\sim30$ BAT-GBM GRBs, we have several bright common events from the activation of a new Soft Gamma Repeater (SGR $0501+4516$), which will provide insight to the lower energy capabilities of GBM. Our cross-calibration results and a catalog of joint GRB spectral fits will be published elsewhere (Stamatikos et al. 2009, in preparation).

Overall, broad-band correlative studies will enable the investigation of GRB spectral and temporal evolution over unprecedented decades of energy, shedding light on the connection between electromagnetic pulse asymmetry, width and spectral softening, while facilitating searches for correlated leptonic emission in context of multi-messenger astronomy \cite{Stamatikos:2009b}. Hence, multi-wavelength work is critical to Fermi's success and benefits the broader astronomical community, by enhancing the science return, without additional demands upon mutual mission resources, thus underscoring Swift's operational and scientific relevancy in the next decade.


\bibliographystyle{aipproc}   

\bibliography{Proc_AIP_Stamatikos_Michael_P_P_astroph}

\IfFileExists{\jobname.bbl}{}
 {\typeout{}
  \typeout{******************************************}
  \typeout{** Please run "bibtex \jobname" to optain}
  \typeout{** the bibliography and then re-run LaTeX}
  \typeout{** twice to fix the references!}
  \typeout{******************************************}
  \typeout{}
 }

\end{document}